\documentclass[%
reprint,
amsmath,amssymb,
aps,
]{revtex4-2}

\usepackage{graphicx}
\usepackage{dcolumn}
\usepackage{bm}
\usepackage[breaklinks=true]{hyperref}
\usepackage{upgreek}
\usepackage{xcolor}

\hypersetup{
	colorlinks   = true, 
	urlcolor     = blue, 
	linkcolor    = blue, 
	citecolor   =  red 
}

\usepackage{tikz}
\usetikzlibrary{calc}
\usepackage[american]{circuitikz}
\usepackage{pgfplots}
\usepgfplotslibrary{fillbetween}
\usetikzlibrary{external}
\usepackage{tikzexternal}
\begin{document}
	
	\preprint{APS/123-QED}
	
	\title{Experimental realization of qubit-state-controlled directional edge states in waveguide QED}%

	\author{Prasanna Pakkiam}
	\affiliation{ARC Centre of Excellence for Engineered Quantum Systems, School of Mathematics and Physics, The University of Queensland, Saint Lucia, Queensland 4072, Australia}
	
	\author{N. Pradeep Kumar}
	\affiliation{ARC Centre of Excellence for Engineered Quantum Systems, School of Mathematics and Physics, The University of Queensland, Saint Lucia, Queensland 4072, Australia}
	
	\author{Chun-Ching Chiu}
	\affiliation{ARC Centre of Excellence for Engineered Quantum Systems, School of Mathematics and Physics, The University of Queensland, Saint Lucia, Queensland 4072, Australia}
	
	\author{David Sommers}
	\affiliation{ARC Centre of Excellence for Engineered Quantum Systems, School of Mathematics and Physics, The University of Queensland, Saint Lucia, Queensland 4072, Australia}
	
	\author{Mikhail Pletyukhov}
	\affiliation{Institute for Theory of Statistical Physics, RWTH Aachen University, 52056 Aachen, Germany}
	
	\author{Arkady Fedorov}
	\affiliation{ARC Centre of Excellence for Engineered Quantum Systems, School of Mathematics and Physics, The University of Queensland, Saint Lucia, Queensland 4072, Australia}
	
	\date{\today}
	
	\begin{abstract}
		We experimentally realise the theoretical proposal for in-situ tunable photonic edge states emerging from qubits coupled to a waveguide with a photonic bandgap. These edge states are directional, exhibiting theoretically zero population in the opposite direction. Our experiment implements a tunable Rice-Mele waveguide configuration, where the directionality of edge states is controlled in-situ by varying the qubit energy. The Rice-Mele waveguide is constructed using lumped resonators coupled to a standard Xmon qubit. We demonstrate the existence of these edge states both actively, via waveguide transmission, and passively, through qubit emission via an edge state. We estimate a $99.4\%$ fidelity in the directionality, constrained by our measurement noise floor. These results hold significant promise for the development of long-range qubit couplers with effectively zero crosstalk.
	\end{abstract}
	
	\maketitle
	
	
	\section{\label{sec:intro}Introduction}
	
	Light-matter interactions that produce directional spontaneous emission of photons have gained significant attention in recent years. Numerous theoretical proposals have been introduced~\cite{Bello2019,Gheeraert2020,Wang2021,Guimond2020,Pakkiam2023}, with several experimental implementations now demonstrating feasibility~\cite{Kim2021,Kannan2023,Joshi2023,Redchenko2023}. The primary interest in developing such light sources stems from the potential they offer for implementing deterministic, long-range quantum communication protocols~\cite{ Cirac1999,Kimble2008,Kurpiers2018,Aziza2024}. Beyond applications in quantum information processing, controlling photon directionality opens up possibilities for realizing novel complex quantum many-body states of photons~\cite{Lodahl2017,Pichler2015,Ramos2016,Mahmoodian2020} and simulation of quantum magnetism~\cite{Bello2019}. 
	
	Chiral atom-photon interfaces have been pioneered in nanophotonic systems, where photon emission directionality depends on photon polarization relative to the waveguide’s propagation mode~\cite{Laucht2012,Lodahl2015,Kiransk2017}. However, the limited emitter-waveguide coupling strength in optical systems presents challenges. In contrast, the microwave domain leverages deep sub-wavelength metamaterials to tailor photon dispersion~\cite{Mirhosseini2018,Zhang2023}. This has led to chiral atom-photon interactions in waveguide QED~\cite{Kim2021} and unidirectional emission via cavity interference by phase manipulation between entangled qubits~\cite{Kannan2023} (based on~\cite{Lodahl2017,Gheeraert2020}). Additionally, in the giant-atom regime, photon directionality arises from the phase difference between two qubit-waveguide coupling points, controlled by modulating coupling strength in time~\cite{Ruostekoski2023}.
	
	Distinct directional photon transport can also be achieved via topological edge states in 1D topological photonic lattices like the Su-Schrieffer-Heeger (SSH) model~\cite{Su1979}. Introducing a quantum emitter in this setting creates atom-photon bound states ~\cite{Liu2016,Scigliuzzo2022,Kumar2023}, breaking the model’s chiral symmetry and resulting in directional edge states. Notably, as shown experimentally in~\cite{Kim2021}, the directionality of these edge states depends on the emitter’s position within the unit cell, making in-situ tuning challenging. In our recent theoretical work~\cite{Pakkiam2023}, we extend this model to the more general Rice-Mele (RM) model~\cite{Rice1982}, demonstrating that edge state directionality can be controlled through tunable parameters, such as the emitter’s frequency. This approach enables in-situ tunability of edge state directionality while preserving the spatial configuration of the emitter.
	
	In this work, we experimentally implement the RM model by suitably engineering a superconducting metamaterial waveguide where each site is represented by a subwavelength lumped element microwave resonator and a transmon (specifically Xmon) qubit coupled to the waveguide acting as a quantum emitter. We observe the formation of photonic bound edge states within the photonic bandgap and investigate their directionality through microwave scattering experiments. Our results show unidirectional photon transport with directionality controlled by the qubit’s frequency. Additionally, time-domain measurements reveal that the photon emitted by qubit can be made to decay both unidirectionally (either toward the left or right port) or bidirectionally by simply tuning the qubit frequency.

	\section{\label{sec:whyRMmodel}Directional edge states in Rice-Mele waveguide }
	
	\tikzsetnextfilename{Fig1}
	\begin{figure}[!ht]
		\includegraphics{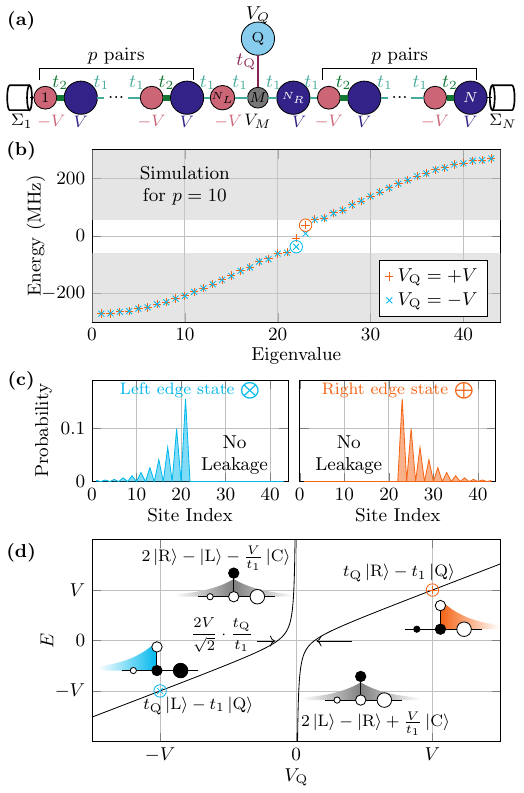}
		\caption{\label{fig:RMmodel} \textbf{Rice-Mele waveguide}. \textbf{(a)} Visual representation of the Rice-Mele waveguide given in Eq.~\ref{eqn:fullHamilIdeal}. The waveguide consists of $N=4p+4$ sites with modulating on-site potentials $-V$ and $+V$ in addition to tunnel-couplings $t_1$ and $t_2$. The qubit is side-coupled with tunnel-coupling $t_\textnormal{Q}$. \textbf{(b)} Numerical simulation of the energy eigenvalues for $p=10$ cells on a Rice-Mele waveguide (taking $V=37.5\,\textnormal{MHz}$, $t_1=120\,\textnormal{MHz}$, $t_2=150\,\textnormal{MHz}$ and $t_\textnormal{Q}=62.5\,\textnormal{MHz}$). The periodicity induces a band-gap. When tuning the qubit energy to $\pm V$, there are two states in the band-gap. \textbf{(c)} The energy eigenstates expressed as the photon probability across the sites along the Rice-Mele waveguide. When tuning the qubit to $-V$ and $+V$, the photon population directs itself perfectly (from the centre) either leftwards or rightwards respectively with zero leakage probability in the opposite direction. The qubit site 44 has been omitted for clarity. \textbf{(d)} The band-gap energy eigenvalues plotted as a function of the qubit energy. The energy eigenstates are typically bidirectional, except when tuning to $V_\textnormal{Q}=\pm V$ where the states are unidirectional.}
	\end{figure}
	
	We begin by constructing a Rice-Mele waveguide out of an array of resonators that act as `sites'. Each site can host a single photon in its harmonic potential well. The sites are tunnel-coupled via a capacitor. This coupling alternates between every pair of sites as $t_1$ and $t_2$ with $t_2>t_1$. The on-site potential offsets (that is, the energy of a photon on that site) also alternate as $+V$ and $-V$. By describing the length of the chain in terms of the number of pairs $p$ strongly coupled together with $t_2$, we can write down the accompanying Hamiltonian across the basis of sites $|m\rangle$:
	
	\begin{align}
		\mathbf{H}_\textnormal{RM}^{n,p}&=\frac{V}{2}\sum_{l=1}^{2p}(-1)^{l}|n+l-1\rangle\langle n+l-1|\nonumber\\
		&-t_2\sum_{l=1}^{p}|n+2l-2\rangle\langle n+2l-1|\nonumber\\
		&-t_1\sum_{l=1}^{p}|n+2l-1\rangle\langle n+2l|.
	\end{align}
	Here, the first line represents the modulating on-site potentials, while the last two lines represent the modulating tunnel couplings. In addition, $n$ is just an indexing offset used when constructing multiple chains. To create edge-states in the SSH model one needs a defect in the periodic chain. A defect like a side-coupled site generates an edge-state with direction dependent solely on the site to which it couples~\cite{Kim2021}. To create directional edge-states, we join two Rice-Mele chains onto a central site with tunnel coupling $t_1$ and have a qubit side-coupled to this central site via tunnel coupling $t_Q$ as shown in Fig.~\ref{eqn:fullHamilIdeal}a~\cite{Pakkiam2023}. The full Hamiltonian is:
	
	\begin{align}\label{eqn:fullHamilIdeal}
		\mathbf{H}_\textnormal{RM} &=
		-t_\textnormal{Q}|M\rangle\langle N|
		+\tfrac{V_\textnormal{Q}}{2}|N+1\rangle\langle N+1| + \tfrac{V_M}{2}|M\rangle\langle M|\nonumber\\
		&-\tfrac{V}{2}|N_L\rangle\langle N_L|+\tfrac{V}{2}|N_R\rangle\langle N_R| - t_1\sum_{m=2p}^{N_R}|m\rangle\langle m+1|\nonumber\\
		&+\mathbf{H}_\textnormal{RM}^{1,p}+\mathbf{H}_\textnormal{RM}^{N_R+1,p} + h.c.,
	\end{align}
	where the Hermitian conjugate applies to all listed terms. The index of the central site is $M=2p+2$, with the adjacent site to the left being $N_L=M-1$, the adjacent site on the right being $N_R=M+1$, while the qubit site is on $Q=4p+4$ (the dimension of this Hamiltonian is $Q$). To break the inversion symmetry between the two Rice-Mele chains, we utilise the modulating on-site potentials $\pm V$ to break the degeneracy between the left and right edge-states. We see this in Fig.~\ref{eqn:fullHamilIdeal}b where we numerically plot the eigenvalues for $p=10$ for qubit energies $V_\textnormal{Q}=\pm V$. In fact, we can manifest an edge-state eigenstate that changes in direction when $V_\textnormal{Q}$ toggles between $+V$ and $-V$ as plotted in Fig.~\ref{eqn:fullHamilIdeal}c. The edge-states stem from the qubit and central site unidirectionally with zero leakage population in the other direction. Fig.~\ref{eqn:fullHamilIdeal}d shows the characteristic anti-crossing observed when sweeping the qubit energy within the band-gap. The eigenstates near the centre are bidirectional in nature, while on the branches with energies $\pm V$, the eigenstates are the edge-states plotted in Fig.~\ref{eqn:fullHamilIdeal}c.
	
	Finally, there is an additional self-energy term that is added for the terminating ports on sites $1$ and $N=3p+3$ with self-energies $\Sigma_1$ and $\Sigma_N$~\cite{mesoDatta,Pakkiam2023}. The imaginary portion of the self-energy is proportional to half the decay rate of a photon on that site into the ports. The decay of photons into the port can be used to probe the Rice-Mele waveguide.
	
	\tikzsetnextfilename{Fig2}
	\begin{figure*}[!ht]
		\includegraphics{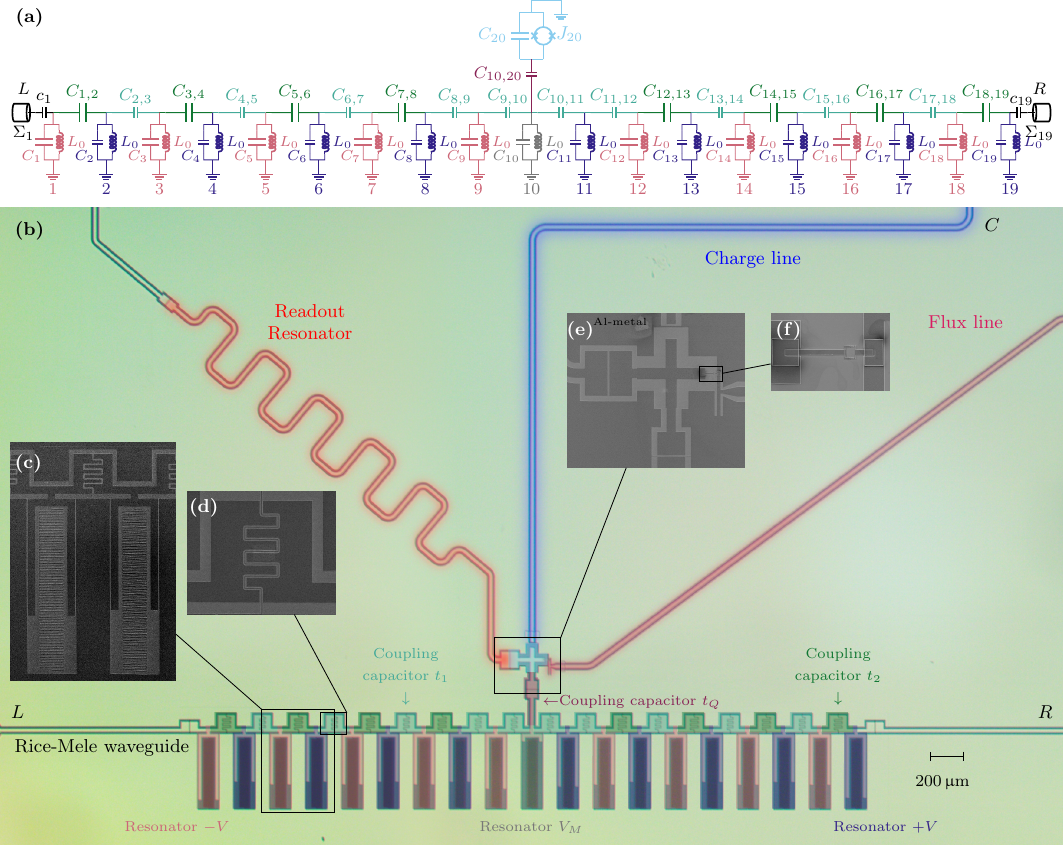}
		\caption{\label{fig:device} \textbf{Main device implemented in cQED}. \textbf{(a)} The equivalent circuit of the Rice-Mele waveguide with $p=4$. The on-site potentials are implemented with lumped LC resonators $C_{i}$ and $L_0$ (site labels are given underneath each LC resonator), while bare lumped capacitors $C_{i,j}$ create the tunnel-couplings. The qubit is formed via a SQUID loop $J_{20}$ shunted by a capacitor $C_{20}$. The qubit side-couples to the waveguide via a capacitor $C_{10,20}$. The ports $L$ and $R$ couple to the waveguide via capacitors $c_i$ and have self-energies $\Sigma_i$. \textbf{(b)} Optical image of the full chip with the zoomed insets showing SEM scans. The device consists of a Rice-Mele waveguide coupled to a Xmon. Each waveguide resonator consists of a lumped capacitance and a meander inductor to ground. The coupling capacitors are formed via two interdigital plates separated by the ground-plane. The Xmon can be independently controlled and read via a charge line and resonator respectively. A flux line is used to control the qubit energy. The Rice-Mele waveguide is addressed via left and right ports $L$ and $R$. \textbf{(c)} SEM image of two resonator sites forming a unit-cell where each site has a meander stripline for the inductor and two prongs arms close to ground for the capacitance to ground. \textbf{(d)} SEM image of interdigital capacitors, bisected by a ground-plane, used to implement $t_1$ and $t_2$. \textbf{(e)} SEM image of the Xmon structure forming the qubit. \textbf{(f)} SEM image of the SQUID loop forming tunable nonlinear element for qubit.}
	\end{figure*}
	
	\tikzsetnextfilename{Fig3}
	\begin{figure*}[!ht]
		\includegraphics{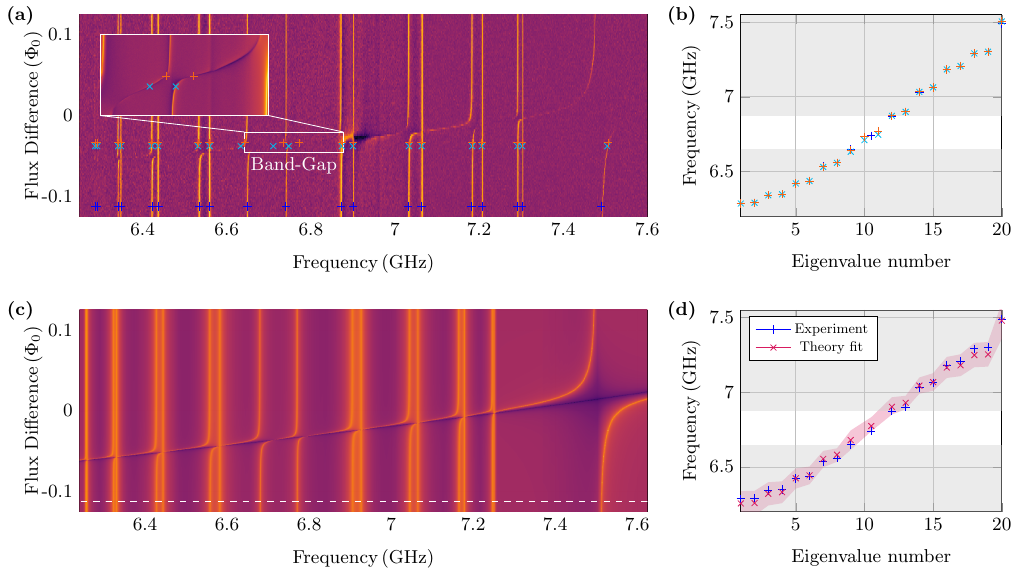}
		\caption{\label{fig:fluxsweep} \textbf{The measured and theoretical transmission spectrum as a function of flux}. \textbf{(a)} The measured transmission spectrum $S_{RL}$. The inset shows the anti-crossing of interest. We swept the flux in the main plot using the global coil, while using the flux line for the inset. Note that the data has been background-subtracted using a median filter for clarity. \textbf{(b)} The spectral peaks are shown across three slices of interest as shown in (a). The non-shaded region in white is the band-gap. \textbf{(c)}-\textbf{(d)} The theoretical best-fit of the transmission response and the associated peak positions. The spectral data was optimised using $t_1=230\pm20~\textnormal{MHz}$, $t_2=280\pm20~\textnormal{MHz}$, $V=40\pm20~\textnormal{MHz}$, $V_M=590\pm50~\textnormal{MHz}$ and $t_q=130\pm20~\textnormal{MHz}$ (for the Hamiltonian given in Eq.~\ref{eqn:fullHamilIdeal}). The fitting bounds shown in red cover the percentiles 2.5 to 97.5.}
	\end{figure*}
	
	To implement this Hamiltonian in cQED, we map each site of the Hamiltonian onto a lumped LC resonator and each tunnel coupling onto a lumped capacitor~\cite{Kim2021,Scigliuzzo2022,Pakkiam2023,Jouanny2024}. Fig.~\ref{fig:device}a shows the equivalent circuit of this mapping. The device consists of a capacitively coupled lumped resonator array forming the Rice-Mele waveguide. The waveguide can be probed via ports $L$ and $R$ (using capacitors $c_i$), which have self-energies $\Sigma_1$ and $\Sigma_{19}$. To modulate the tunnel-couplings we modulate the interlinking capacitances $C_{i,j}$. To modulate the on-site couplings $\pm V$, we use uniform resonator inductances $L_0$, while modulating the resonator capacitances $C_i$ to ground. We create the qubit (side-coupled via the capacitor $C_{10,20}$) using an Xmon formed via a shunting capacitor $C_{20}$ and a SQUID (superconducting quantum interference device) loop $J_{20}$ composed of two Josephson junctions. The SQUID loop enables us to tune the qubit energy by changing the magnetic field penetrating the loop.
	
	All elements in the circuit are superconducting elements fabricated with conventional electron-beam lithography (EBL). First, a polymeric resist mask with the lumped-element patterns was transferred onto the silicon substrate by EBL. A layer of aluminium film was deposited onto the silicon substrate using electron-beam evaporation and the resist mask was lifted off via dimethyl sulfoxide (DMSO). The SQUID loop with two Josephson junctions, which serves as a non-linear inductor for the side-coupled qubit, was fabricated in a similar manner except using shadow evaporation. 
	
	Fig.~\ref{fig:device}b shows an optical image (with zoomed scanning electron microscope images in the insets) of the overall structure of the fabricated device. The lighter region is aluminium while the darker regions, in the gaps, are bare silicon. The insets show SEM (scanning electron microscope) scans in which the darker regions are aluminium while the ligher regions are bare silicon. The zoomed SEM scan of a unit-cell of two resonators is shown in Fig.~\ref{fig:device}c. The resonators used to implement the $\pm V$ couplings are formed via uniform meander inductors with $C_i$ controlled by geometrically varying by the length of two prongs close to ground. For example, when the prongs are longer, the capacitance $C_i$ is larger, leading to a lower resonant frequency; thereby implementing a $-V$ potential. A zoomed SEM scan of the interlinking capacitors (used to implement the $t_1$ and $t_2$ couplings) in Fig.~\ref{fig:device}d shows the interdigital structure bisected by a ground line to ensure that there are no breaks in the ground plane~\cite{Kim2021}. We employ the same principle of using a bisecting ground plane in forming the capacitors connecting to the ports and the qubit. Fig.~\ref{fig:device}e shows a zoomed SEM scan of the qubit which is an Xmon that is side-coupled (EXPLAIN) to the central-site of the array. The qubit energy $V_\textnormal{Q}$ can be independently tuned by changing the magnetic flux entering the SQUID loop shown in SEM scan in Fig.~\ref{fig:device}f. We placed the SQUID loop close to the qubit flux line to ensure sufficient flux tunability. We provide an additional flux offset via a global coil placed underneath the chip to ensure maximal flexibility. Finally, the Xmon can be controlled and read out independently via the charge line (port $C$) and readout resonator as highlighted in blue and red respectively in Fig.~\ref{fig:device}b.  To measure this device, we placed it in a \emph{BluFors LD} dilution refrigerator with a mixing chamber temperature of approximately 7\,mK.
	
	\section{Probing directional edge states}
	\subsection{Directional photon scattering}
	We first probed the Rice-Mele waveguide in transmission over the ports $L$ and $R$ using a vector network analyser (VNA). When varying the flux in the global coil, the resulting $S_{RL}$ plotted in Fig.~\ref{fig:fluxsweep}a shows the expected 19 peaks (for each resonator in the RM-waveguide). Since the qubit is side-coupled, the qubit peak will be a transmission dip as the signal is absorbed and thus, diverted away from the main transmission path. However, this dip is typically not visible within the noise floor of the measurement. However, the qubit interaction manifests via anti-crossings. When mapping the transmission peaks with the qubit far-detuned, we get the blue crosses plotted in Fig.~\ref{fig:fluxsweep}b. 
	
	\tikzsetnextfilename{Fig4}
	\begin{figure}[!htb]
		\includegraphics{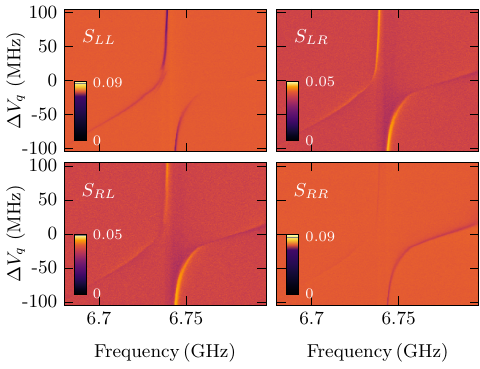}
		\caption{\label{fig:figVNAzoom} \textbf{Scattering matrix of the band-gap modes as a function of flux}. The anti-crossing shown in Fig.~\ref{fig:fluxsweep}a was probed via transmission and reflection on both waveguide ports $L$ and $R$ using a VNA. We swept the flux using the flux line. The data was background-subtracted via a median-filter for clarity.}
	\end{figure}

	There is a clear separation of two clusters of peaks converging onto a so-called band-gap in the centre housing a single state. The photonic wavefunction eigenstates for the peaks in the shaded regions spread over the entire waveguide to overlap with the ports $L$ and $R$; thus, they are akin to a conduction band with propagating modes. The state that lives in the band-gap is localised and forms a static evanescent mode. Given that we have a moderate number of nodes, the emergence of the band-gap appears less pronounced. In order to unambiguously distinguish the peaks inside and outside the band-gap, we note the specific characteristic pattern of the waveguide modes to which the qubit interacts and forms an anti-crossing. Due to the even/odd Bloch solutions that have zero population on the central site $M$, the qubit only interacts with alternate waveguide modes when outside the band-gap. Within the band-gap there is a characteristic pattern where three adjacent modes interact with the qubit mode to form anti-crossings.
	
	When sweeping the qubit energy through the band-gap as shown in the inset of Fig.~\ref{fig:fluxsweep}a, we can access the working points, shown in light blue and orange, that manifest in the leftward and rightward edge states respectively. We fitted the locations of the far-detuned transmission peaks to our Hamiltonian model and then separately fitted the size of the anti-crossings. From this, we obtained $t_1=230\pm20~\textnormal{MHz}$, $t_2=280\pm20~\textnormal{MHz}$, $V=40\pm20~\textnormal{MHz}$, $V_C=590\pm50~\textnormal{MHz}$ and $t_q=130\pm20~\textnormal{MHz}$ (where the error bars come from the standard deviations of the distributions when taking a random 10000 samples over the bootstrapping method). When calculating the local density of states~\cite{mesoDatta, Pakkiam2023} using a coupling of 18\,MHz (to match the port-couplings inferred from the later measurements), we get the matching theoretical transmission in Fig.~\ref{fig:fluxsweep}c. The theoretical best-fit of the far-detuned spectrum is shown in Fig.~\ref{fig:fluxsweep}d. The shape of the spectrum agrees with our Hamiltonian model to within uncertainty bounds. The large anti-crossing to the right of the spectrum is due to the large $V_M$ that places the central site outside the waveguide modes. Thus, in the band-gap the qubit only interacts with sites $N_L$ and $N_R$ to give an effective 3-site model~\cite{Pakkiam2023}.
	
	\tikzsetnextfilename{Fig5}
	\begin{figure}[!htb]
		\includegraphics{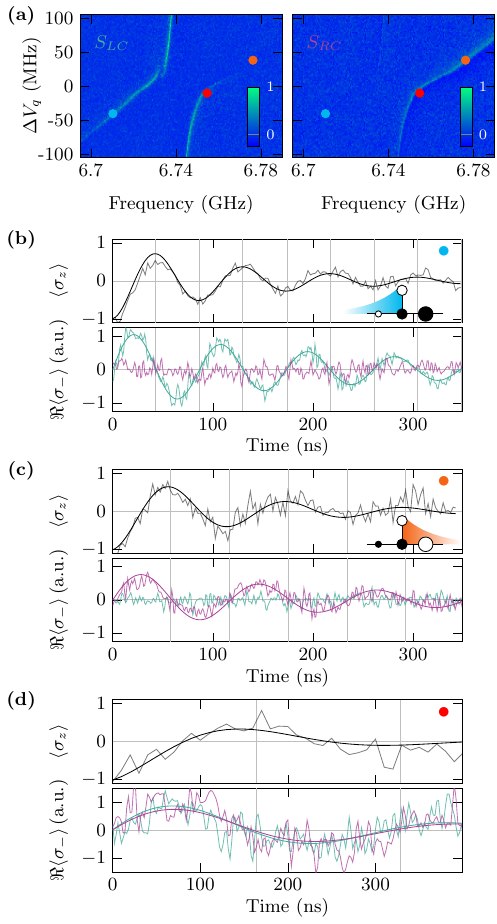}
		\caption{\label{fig:figtimeDomain} \textbf{Probing edge-states with qubit}. \textbf{(a)} The anti-crossing was measured in transmission (a.u.) via the charge-line $C$ with the responses measured onto the ports $L$ and $R$. There is an asymmetry in the response similar to the reflection plots in Fig.~\ref{fig:figVNAzoom}. We swept the flux using the flux line. \textbf{(b)}-\textbf{(d)} The response of the qubit when measured at the three points marked in (a), with (b) being the leftward edge-state $V_\textnormal{Q}=-V$ coupling only to port $L$, (c) being the rightward edge-state $V_\textnormal{Q}=V$ coupling only to port $R$ and (d) being a bidirectional state coupling equally to both ports. We peformed a typical Rabi experiment: initialising qubit into ground state, waiting for some time and then measuring the population. The $\langle\sigma_z\rangle$ plots were taken with the qubit readout resonator, while the $\Re\langle\sigma_-\rangle$ measurements were done by simply measuring the microwave signal on ports $L$ and $R$. The repetition count in the ensemble average was 65536 for (b) and (c) and 8192 for (d). We post-processed the data with a 75\,MHz low-pass filter for clarity.}
	\end{figure}

	To probe the directionality, we then measured the full scattering matrix in Fig.~\ref{fig:figVNAzoom}, using a VNA, around the central anti-crossing where the qubit interacts with the band-gap state. The transmissions (off-diagonal entries $S_{LR}$ and $S_{RL}$) are symmetric as expected, although due to the noise floor, the zeroing of the qubit peak is not visible. However, when probing the ports via reflection (the diagonal entries $S_{LL}$ and $S_{RR}$), there is a clear asymmetry. That is, the upper anti-crossing band only shows full visibility when probed via port $R$ and similarly the lower anti-crossing band is clearly visible only when probed via port $L$. This shows the directionality of the edge-state. That is, because it only points exclusively in a single direction, it only couples to one of the ports.

	We then probed the ability of the qubit to emit a photon to only one of the ports on demand by utilising the edge-states found in the band-gap. To control the qubit via the charge line, we used a typical IQ-modulation setup. We first probed the anti-crossing by measuring transmission via the charge line. Fig.~\ref{fig:figtimeDomain}a shows the transmission from the charge line $C$ onto the ports $L$ and $R$. We once again see the same asymmetry seen in the reflection plots in Fig.~\ref{fig:figVNAzoom}, where the upper and lower anti-crossing bands are only fully visible when measuring on ports $L$ and $R$ respectively.
	
	\subsection{Directional photon emission from qubit}
	We first positioned ourselves onto the point $V_\textnormal{Q}=-V$ (blue point indicated in Fig.~\ref{fig:figtimeDomain}a). We performed a typical Rabi experiment in which we first initialised the qubit into the ground state, rotated it about the $x$-axis by keeping the drive on over a range of wait-times and then finally measuring the population probability. Fig.~\ref{fig:figtimeDomain}b shows a typical Rabi curve as probed with the qubit readout resonator. As usual, this measures the expectation over the computational basis: $\langle\sigma_z\rangle$. We then performed the exact same protocol but for the measurement, we instead just observed the ports $L$ and $R$ for emitted photons (the readout readout probe signal on the qubit resonator remained switched off). When taking the ensemble average at a time-point after the qubit drive finishes, we see a similar sinusoidal curve (but only on port $L$). Since the process that couples the qubit to the port is a $\sigma_-$ decay, the ports effectively measure $\Re\langle\sigma_-\rangle$. Since we do heterodyne detection we measure the full $\sigma_-$, but we can choose to plot one real part. Since $\Re\langle\sigma_-\rangle$ is only sensitive to states on the $xy$-plane, we observe that the signal's peaks and troughs occur when $\langle\sigma_z\rangle=0$. Note that we extracted the $\Re\langle\sigma_-\rangle$ by taking the quadrature angle that maximises the signal on the IQ-plane~\cite{Bozyigit2010}. The resulting sinusoidal response is only observed on port $L$ as the edge-state is leftward facing. We switched the direction of the unidirectional emission by moving to $V_\textnormal{Q}=V$ (orange point indicated in Fig.~\ref{fig:figtimeDomain}a). Once again, we observe a sinusoidal response in $\Re\langle\sigma_-\rangle$ that is $\pi/2$ out of phase with the $\langle\sigma_z\rangle$ as shown in Fig.~\ref{fig:figtimeDomain}c. As before, the response is only seen on one of the ports, specifically port $R$. From the signal strengths of the edge-state sinusoids in Fig.~\ref{fig:figtimeDomain}b and Fig.~\ref{fig:figtimeDomain}c, we estimate the average directionality (the ratio of population spread in the intended direction to the opposite direction) to be at least 22\,dB given our measurement noise floor as detailed in Appendix~\ref{sec:appen:chicalc}. Equivalently stated, the fidelity (ratio of the signal in the intended direction over the total signal onto both ports) was at least $99.4\%$.
	
	Finally, we moved to a point where there was signal seen in both ports indicated as red point in Fig.~\ref{fig:figtimeDomain}a. Fig.~\ref{fig:figtimeDomain}d shows the response when moving to a point where there is theoretically an equal proportion of leftward and rightward facing edge states; that is, a bidirectional state. Since, the qubit proportion is lesser in this superposition, the smaller effective drive coupling led to a slower Rabi oscillation. Another interesting observation was that the decay measured in all three $\langle\sigma_z\rangle$ signals ($130\pm10\,\textnormal{ns}$, $130\pm10\,\textnormal{ns}$ and $140\pm20\,\textnormal{ns}$), were approximately two times faster than that measured in $\Re\langle\sigma_-\rangle$ ($260\pm50\,\textnormal{ns}$, $250\pm40\,\textnormal{ns}$, $260\pm40\,\textnormal{ns}$ and $260\pm40\,\textnormal{ns}$). This is consistent with the trajectory of a $T_1$-limited process.

	We then measured the decay of the photons onto the ports. We initialised the qubits onto the $xy$-plane, the apex of the sinusoidal $\Re\langle\sigma_-\rangle$ plots in Fig.~\ref{fig:figtimeDomain}, and then observed the state over time. On taking the ensemble average of all the time-traces, we get an approximate exponential decay. This decay of $110\pm20\,\textnormal{ns}$ matches the decay-time in the $\langle\sigma_z\rangle$ measurements to within uncertainty. The decay can be used to numerically infer the self-energy of the ports by inspecting the imaginary part of the eigenvalues on adding the self-energy terms to the fitted Hamiltonian~\cite{Pakkiam2023}. We found that the port self-energies were approximately $\Sigma_1=\Sigma_2=-j18\,\textnormal{MHz}$.
	
	\tikzsetnextfilename{Fig6}
	\begin{figure}[!ht]
		\includegraphics{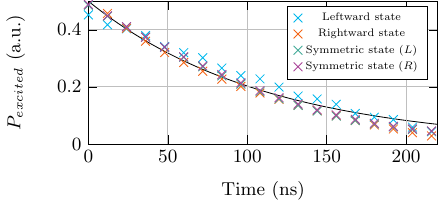}
		\caption{\label{fig:figExpDecays} Qubit state decay as measured directly with the ports. Here the qubit was excited via a $X_{\pi/2}$ pulse onto the $xy$-plane. Then it was continously monitored over time. The points shown here are taken from an ensemble average of many time traces (65536, 65536, 8192 and 8192 respectively).}
	\end{figure}
	
	\section{Conclusion}
	
	In this work we implemented a Rice-Mele waveguide that can host edge states with its direction in-situ tunable via the energy of a qubit as given in our previous proposal~\cite{Pakkiam2023}. We first verified the presence of unidirectional edge-states both spectrally via transmission and reflection. We then verified that the qubit can be made to emit photons exclusively to a given port via said edge-states to further confirm the unidirectionality. Future pathways could include direct single-shot photon measurements by attaching TWPAs to both the ports $L$ and $R$. Another extension is to utilise the emitted photons to couple multiple qubits along the Rice-Mele waveguide. By toggling the direction of the edge-states, one can demonstrate finite coupling (when the edge-states of adjacent qubits face one another) and zero coupling (when the edge-states of adjacent qubits face away from one another) to significantly reduce crosstalk~\cite{Kim2021}. Finally, the waveguide can be made significantly smaller by utilising spiral inductors~\cite{Fink2020}, superconductors with high kinetic inductance and parallel-plate capacitors\cite{McFadden2024}.
	
	\begin{acknowledgments}
		The authors were supported by the Australian Research Council Centre of Excellence for Engineered Quantum Systems (EQUS, CE170100009).
	\end{acknowledgments}
	
	\appendix

	\section{Experimental apparatus}\label{sec:appen:exptApparatus}
	
\begin{figure*}[!ht]
		\includegraphics{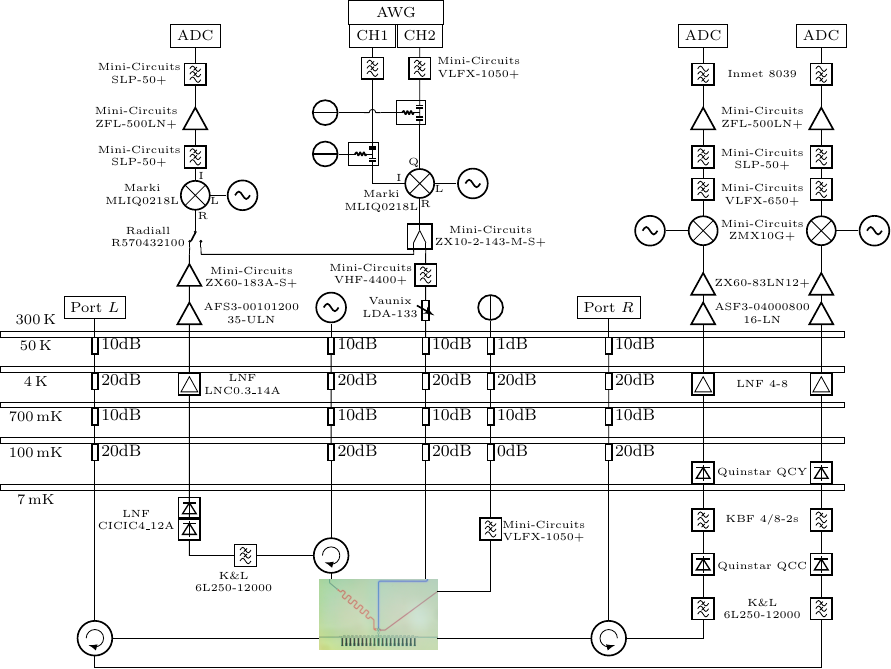}
		\caption{\label{fig:figExptApparatus} \textbf{The wiring diagram of experiment}. The wiring diagram shows the wiring within the different stages of the BluFors LD fridge as well as the room temperature apparatuses. Note that for the transmission measurements between ports $L$ and $R$, the inlets Port $L$ and Port $R$ were connected to a VNA along with return lines after the second amplifier in the room temperature portion. The diagram shown here is for the time-domain measurements. The 2GSPS AWG and 2.5GSPS ADC ports are from the \emph{Tabor P2584M} PXI transceiver. Since this transceiver only has 2 ADC ports, we had a switch (not shown) to toggle readout-resonator and port measurements. The microwave sources are all \emph{R\&S SGS100A} models, while the DC sources are all ports from a single \emph{Stahl BSA-16-40} unit. The circulator on the readout resonator line is \emph{Raditek RADC-4-8-Cryo-0.02-4K-S23-1WR-B}, while the remaining two circulators are \emph{Quinstar QCN}. The Bias-Ts shown attenuate the DC component by 20\,dB. The switch shown in the circuit is used to aid in mixer calibration of the IQ-mixer used in generating the signal for the qubit charge line.}
	\end{figure*}

	The full fridge and measurement circuits are shown in Fig.~\ref{fig:figExptApparatus}. The fabricated chip was bonded to a PCB placed on the mixing chamber inside a copper casing, encased in copper and mumetal for proper thermalisation, radiation shielding and magnetic shielding. The lab instrumentation was operated via our open source toolkit \emph{SQDToolz} available on GitHub~\cite{sqdtoolz}.
	
	\tikzsetnextfilename{Fig7}

	\section{Driving the transitions}\label{sec:appen:driveRabisRamseys}
	
\begin{figure*}[!ht]
		\includegraphics{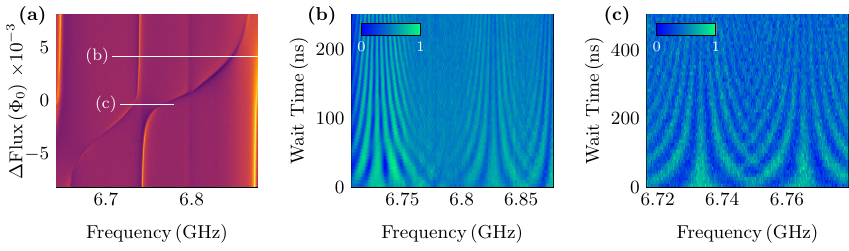}
		\caption{\label{fig:figFancyRamsey} \textbf{Detuned Ramsey plots around spectral peaks}. $\mathbf{(a)}$ The anti-crossing of the band-gap states as a function of frequency and flux measured in transmission from ports $L$ to $R$ as in Fig.~\ref{fig:fluxsweep}. \textbf{(b)}-\textbf{(c)} The associated Ramsey plots for the line-cuts shown in (a). The colours signify the excited qubit state probability as a function of the Ramsey detuning frequency, along the white line in (a), and the Ramsey wait-time.}
	\end{figure*}

	We characterised the qubit using a simple Rabi sequence where we initialise the qubit into the ground state, drive for a specified period of time and then measuring the final state. We were able to drive the qubit directly by setting the flux and frequency to different parts of the energy landscape as shown in Fig.~\ref{fig:figtimeDomain}. To show the validity of this direct drive, we ran a Ramsey sequence over the spectral lines. We first found a spectral line and ran a Rabi cycle to calibrate the $X_{\pi/2}$ gate. The Ramsey protocol consisted of a $X_{\pi/2}$ gate to tip the initial ground state onto the $xy$-plane. Then we waited over a series of times before untipping the state via another $X_{\pi/2}$ gate. When detuning away from the spectral lines, we see oscillations due to off-resonant $z$-rotations. Consider the band-gap anti-crossing in Fig.~\ref{fig:figFancyRamsey}a. We now set the flux across two line cuts upon which we perform the detuned Ramsey protocol. The corresponding Ramsey plots as a function of detuning and wait-time are shown in Fig.~\ref{fig:figFancyRamsey}b and Fig.~\ref{fig:figFancyRamsey}c. This demonstrates that we can coherently drive the transitions directly using the charge line as required in the time-domain plots shown in Fig.~\ref{fig:figtimeDomain}.
	
	\tikzsetnextfilename{Fig8}
	
	\section{Approximating directionality}\label{sec:appen:chicalc}
	
	The definition of $\chi$ is the ratio of total populations spread in the intended direction versus that in the opposite direction in the waveguide with respect to position $M$~\cite{Pakkiam2023}. We can estimate that from the signal amplitudes at the Rabi frequencies measured on different ports in Fig.~\ref{fig:figtimeDomain} to estimate this ratio. However, the amplification and line attenuations on the two paths from the $L$ and $R$ may not necessarily be uniform. That is:
	
	\begin{align}
		\chi_l&=\frac{s_{l,L}}{s_{l,R}}\cdot\frac{g_{L}}{g_{R}}\\
		\chi_r&=\frac{s_{r,R}}{s_{r,L}}\cdot\frac{g_{R}}{g_{L}}
	\end{align}
	where $\chi_l$ and $\chi_r$ are the directionalities of the rightward and leftward edge-states given the signal $s_{m,n}$ being the measured signal amplitude of edge-state $m$ on port $n$. The conversion gains on the ports $L$ and $R$ are $g_L$ and $g_R$ respectively. Thus, to cancel out the conversion gains, one can approximate an average directionality as:
	
	\begin{equation}
		\chi\approx\sqrt{\chi_l\chi_r}.
	\end{equation}
	To calculate the signal amplitude, we multiplied the signal with the sinusoid at the Rabi frequency, low-pass filtered it to $6\,\textnormal{MHz}$ (to obtain the demodulated zero-frequency component) and then integrated it to obtain the final signal strength. Note that the ratios cancel out differences in integration due different number of periods. Given the short sampling period and noise, we ran a bootstrapping algorithm to randomly sample the signal (interpolating the resulting signal using nearest-sampling) before extracting the signal amplitude. We ran 10000 samples to get a distribution of the signal strengths (a.u.): $s_{l,L}=108.2\pm0.3$, $s_{l,R}=0.3\pm0.3$, $s_{r,L}=0.7\pm0.3$, $s_{r,R}=54.5\pm0.3$. This gives approximately $\chi\approx167.5$, which is approximately 22\,dB. This figure can be re-expressed as a fidelity metric defined as the proportion of the directed signal on the desired port over the total signal on both ports. It can be shown that the fidelity is thus, simply $\chi/(1+\chi)$, to give $99.4\%$.
	
	\bibliography{apssamp}
	
\end{document}